\documentclass[aps,prd,preprint,tightenlines,superscriptaddress,showpacs]{revtex4}
\usepackage{graphicx}
\usepackage{epsfig}
\usepackage{dcolumn}
\usepackage{bm}
\usepackage{color}

\newcommand{\x}{X(3872)}
\newcommand{\y}{Y(4260)}
\newcommand{\z}{Z_c(3900)}

\newcommand{\yone}{Y(4360)}
\newcommand{\ytwo}{Y(4660)}
\newcommand{\lum}{{\cal L}}

\newcommand{\BR}{{\cal B}}

\newcommand{\pip}{\pi^+}
\newcommand{\pim}{\pi^-}

\newcommand{\psip}{\psi(2S)}

\newcommand{\EE}{e^+e^-}

\newcommand{\pp}{\pi^+\pi^-}
\newcommand{\kk}{K^+K^-}

\newcommand{\ppjpsi}{\pi^+\pi^- J/\psi}
\newcommand{\pppsip}{\pi^+\pi^- \psip}
\newcommand{\Lam}{\Lambda_c^+\Lambda_c^-}

\def\PRD{Phys. Rev. D}

\def\EPJC{Eur. Phys. J. C}

\begin{document}
	
\preprint{}
\title{Exploring the $e^+e^-\to \Lambda_c^+\Lambda_c^-$ cross sections}

\author{Yong~Xie}
\affiliation{Institute of Frontier and Interdisciplinary Science, Shandong University, Qingdao 266237}
\author{Zhiqing~Liu}
\email{z.liu@sdu.edu.cn}
\affiliation{Institute of Frontier and Interdisciplinary Science, Shandong University, Qingdao 266237}

\date{\today}

\begin{abstract}
A simultaneous fit is performed to the $\EE\to\Lam$ cross section data measured by Belle and BESIII 
from threshold up to 5.4~GeV. In order to accommodate both the BESIII measurement near threshold
and the Belle observation of a resonance $Y(4630)$, we build a composite PDF with a Breit-Wigner 
resonance and a continuum contribution to model the full cross section line shape of $\EE\to\Lam$.
The fit gives a mass of $M=[4636.1_{-7.2}^{+9.8} ($stat$)\pm 8.0($syst$)]$~MeV/$c^2$, a width of 
$\Gamma_{\rm tot}=[34.5_{-16.2}^{+21.0} ($stat$)\pm 5.6($syst$)]$~MeV, and
$\Gamma_{e^+ e^-}\BR(Y(4630)\to\Lambda_c^+\Lambda_c^- )=[18.3_{-6.1}^{+8.8} ($stat$)\pm 1.1($syst$)]$~eV/$c^2$
for the resonance. The width of $Y(4630)$ from our study is narrower than the previous Belle fit.
The mass and width of $Y(4630)$ also show good agreement with 
a vector resonance $Y(4626)$ recently observed in $D_s^+D_{s1}(2536)^-$ by Belle. 
\end{abstract}

\maketitle
	
\section{Introduction}
\label{intro}

Over the last decade, many charmoniumlike particles was observed in experiment, such
as the $\x$~\cite{x3872}, $\y$~\cite{y4260}, and $\z$~\cite{zc3900}. These new particles give us strong indication that they
are non-standard $q\bar{q}$ mesons from the original quark model, and are good candidates for the exotic hadron states 
allowed by QCD~\cite{xyz-review}. Among them, the $\y$ state was firstly observed by the {\it BABAR} experiment in 
initial-state-radiation (ISR) process $\EE\to\gamma_{ISR}\ppjpsi$~\cite{y4260}, and thus has a quantum number $J^{PC}=1^{--}$
(also called vector state). Later on, more vector states were observed in experiment, both in ISR process ($\yone$ and 
$\ytwo$ in $\EE\to\gamma_{ISR}\pppsip$~\cite{y4360}) and in direct $\EE$ annihilation ($Y(4220)$ in $\EE\to\omega\chi_{c0}/
\pip\pim h_c/\pip D^0D^{*-}$~\cite{wcc0,pphc,piddstar}). All these vector $Y$-states, together with the $\psi(4040)$, $\psi(4160)$,
and $\psi(4415)$~\cite{PDG} obviously make an overpopulation of the vector charmonium states above 4~GeV in the 
potential model~\cite{potential}, and suggest some of them could be exotic hadron states.

To understand the nature of these newly observed vectors, the OZI-allowed decay, i.e. final state involving charm mesons 
or charm baryons is interesting and important. The Belle experiment for the first time studied the charmed-baryon 
pair production via ISR, and measured the exclusive cross section $\sigma(\EE\to\Lam)$ from threshold up to 
5.4~GeV~\cite{belle-lamc}. An enhancement near the production threshold is observed, and Belle conclude it's
a new vector resonance $Y(4630)$, with mass $M=[4634_{-7}^{+8} ({\rm stat})_{-8}^{+5} ({\rm syst})]$ MeV/$c^2$ and width 
$\Gamma= [92_{-24}^{+40} ({\rm stat})_{-21}^{+10} ({\rm syst})]$ MeV. The observation of $Y(4630)$ immediately attracted
people's interest, and is explained as a baryonium~\cite{baryonium}, or a tetraquark state~\cite{tetra}. The BESIII
experiment also studied the $\EE\to\Lam$ process. Using scan data sets from 4.575~GeV to 4.600~GeV, BESIII was able
to achieve a precise $\EE\to\Lam$ cross section measurement near threshold~\cite{bes3-lamc}. Unlike a simple
resonance line shape, the BESIII measurement shows a fast-rise nonzero production cross section near the $\Lam$ 
threshold, and then followed by a plateau up to 4.600~GeV. This behavior challenges the $Y(4630)$ interpretation for the
$\EE\to\Lam$ cross section near threshold, and hints us there is more complicated dynamics behind in the charmed baryon pair 
production process.

Traditionally, the production cross section of $\EE\to B\bar{B}$ (here $B$ denotes a spin-1/2 baryon) is calculated assuming
one photon exchange dominates the interaction, and is parameterized in terms of electro-magnetic form factors
$\sigma_{B\bar{B}}(s)=\frac{4\pi\alpha^2C\beta}{3s}|G_M|^2(1+\frac{2m_B^2}{s}|\frac{G_E}{G_M}|^2)$~\cite{xsec-bbar},
where $G_E$, $G_M$ is the electro and magnetic form factors, $\beta$ is the baryon velocity, and $C$ is the
Coulomb factor. The BESIII measurement has tested such kind of parameterization in $\EE\to\Lam$, and found it can not 
describe the cross section near the $\Lam$ threshold~\cite{bes3-lamc}. Similar phenomenon has also been observed in the
$\EE\to p\bar{p}$~\cite{babar-pp,bes3-pp} and $\EE\to\Lambda\bar{\Lambda}$~\cite{bes3-lamlam} processes.
Recently, the CMD-3 experiment has done a fine scan near the $p\bar{p}$ threshold, and confirms the fast-rise structure
happens within about 1~MeV starting from the $p\bar{p}$ threshold~\cite{cmd-3}. To demonstrate this fast
variation in the cross section line shape, an exponential saturated function is used to parameterized the
$\EE\to p\bar{p}$ process near threshold. What's more, CMD-3 found the exponential saturated function
can also be used to describe the jump in the $\EE\to3(\pp),~\kk\pp$ cross sections.

Inspired by these observations, we reinterprete the $\EE\to\Lam$ cross section data with a new parameterization
method. In this paper, a combined fit to the $e^+e^-\to\Lambda_c^+\Lambda_c^-$ cross sections measured 
by the Belle and BESIII experiments is performed. We aim to improve the fit to the cross section line shape near 
threshold and obtain a better estimation of the $Y(4630)$ resonant parameters.

\section{data}

\begin{figure}
\begin{center}
\includegraphics[width=12cm]{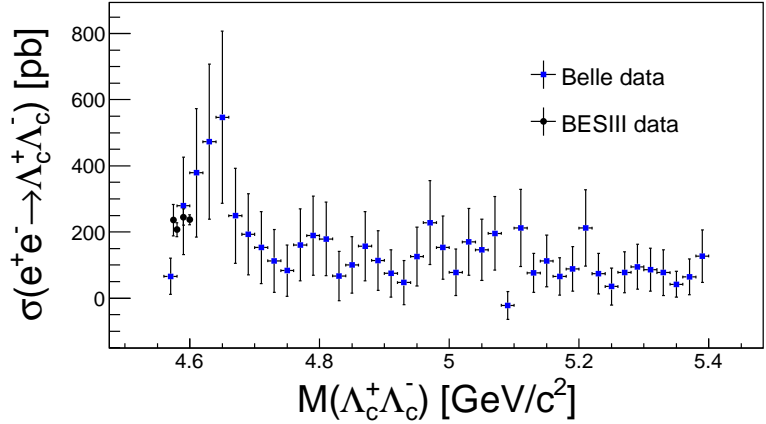}
\caption{The cross sections of $e^+e^-\to\Lambda_c^+\Lambda_c^-$ measured by Belle (blue squares with error bars) 
and BESIII (black dots with error bars).}
\label{xsec-data}
\end{center}
\end{figure}

The data used in this study comes from both the Belle experiment and the BESIII experiment. The Belle
experiment studied the $\EE\to\Lam$ process using ISR method with an integrated luminosity of 695~fb$^{-1}$
on or near the $\Upsilon(4S)$ resonance~\cite{belle-lamc}, and reported the dressed cross section 
$\sigma(\EE\to\Lam)=\frac{dN/dm}{\eta^{\rm tot} d\mathcal{L}^{\rm int}/dm}$, where $d\mathcal{L^{\rm int}}/dm$ is the differential
ISR luminosity, $\eta^{\rm tot}$ is the total efficiency (including branch fractions), 
and $dN/dm$ is the differential mass spectrum~\cite{xsec-formula}.
The BESIII experiment analyzed the scan data sets at $\sqrt{s}=4574.5, ~4580.0, ~4590.0$ and $4599.5$~MeV
~\cite{bes3-lamc}, and reported the born cross section $\sigma(\EE\to\Lam)=\frac{N}{\epsilon\mathcal{L}^{\rm int}f_{VP}f_{ISR}\BR}$,
where $N$ is the number of signal events, $\mathcal{L}^{\rm int}$ is the integrated luminosity, $\epsilon$ is the detection
efficiency, $\BR$ is the branch fraction, $f_{ISR}$ is the radiative correction factor, and $f_{VP}$ is the vacuum
polarization factor. To achieve a fair comparison between Belle and BESIII cross section data, and to follow the
convention vacuum polarization is usually absorbed into the $\Gamma_{ee}$ of a resonance, we convert
the BESIII born cross section into dressed cross section by multiply $f_{VP}=1.055$. Figure~\ref{xsec-data} 
shows the cross section data both from Belle and BESIII measurements.

\section{Fit to data}

\subsection{Fit the $\sigma(\EE\to p\bar{p})$ data}
\begin{figure}
\begin{center}
\includegraphics[width=8cm]{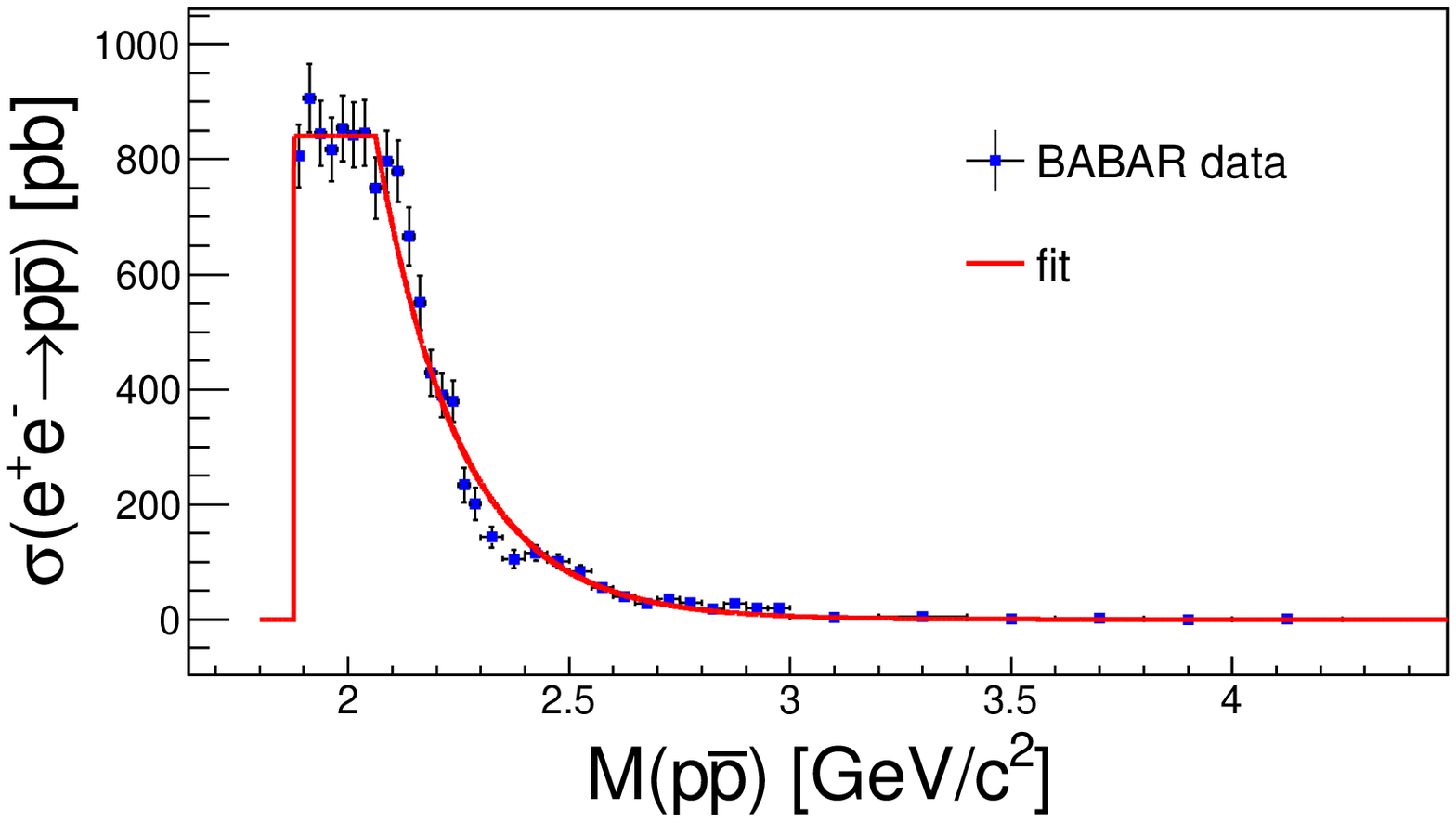}
\includegraphics[width=8cm]{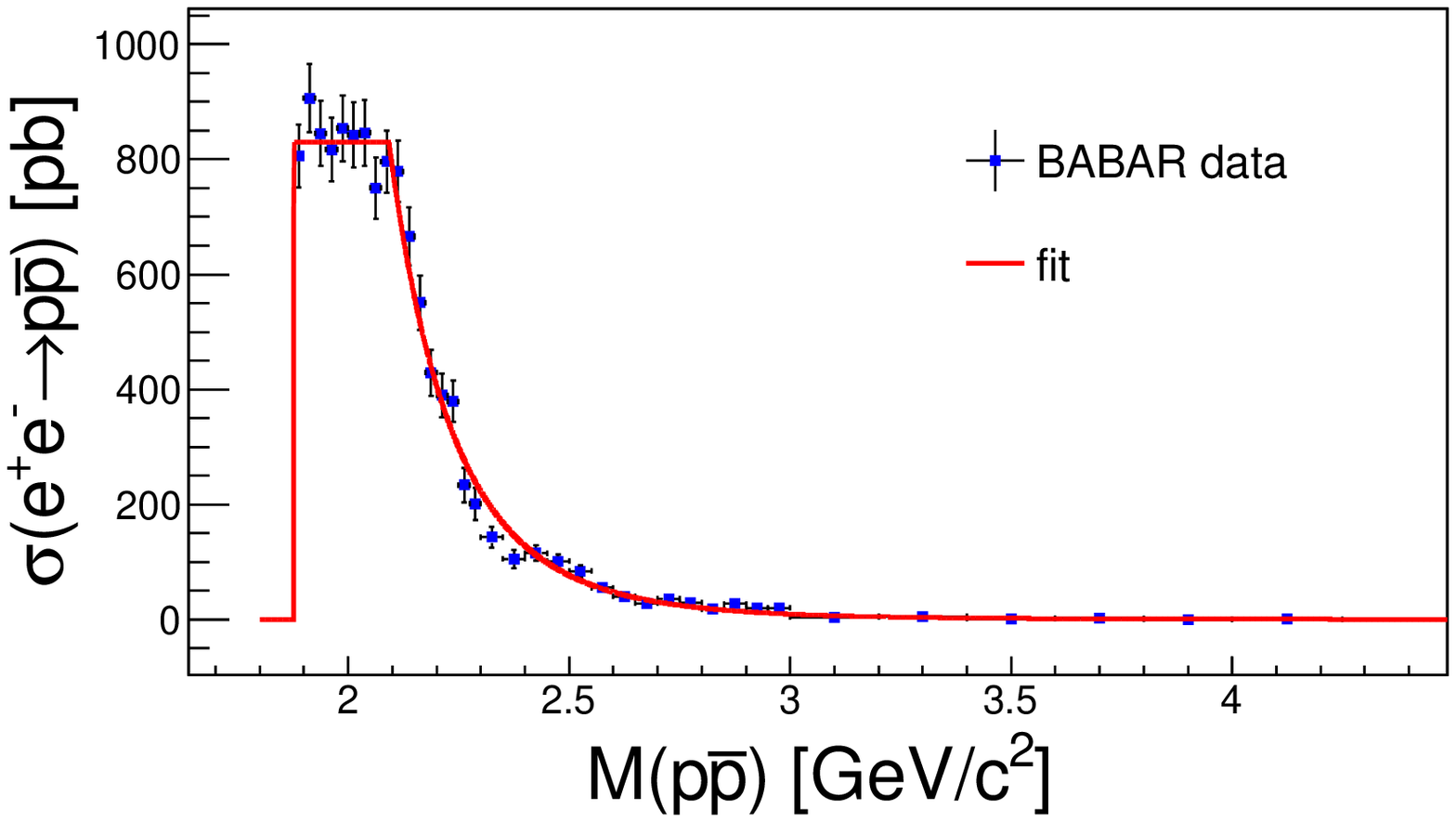}
\caption{The cross section data of $e^+e^-\to p\bar{p}$ measured by {\it BABAR} Collaboration, and a fit
to the cross section with an exponential saturated function and an exponential damping (the left panel)
or a power-law damping (the right panel) at higher energy, respectively.
}
\label{xsec-pp}
\end{center}
\end{figure}

In order to describe the $\EE\to\Lam$ cross section line shape, we first investigate the $\EE\to p\bar{p}$ process, which has
been widely studied by $BABAR$~\cite{babar-pp}, BESIII~\cite{bes3-pp} and CMD-3~\cite{cmd-3} Collaborations.
Figure~\ref{xsec-pp} shows the $\EE\to p\bar{p}$ cross section data measured by $BABAR$ in a wide range. A fast-rise
structure near threshold was observed, which is consistent with the CMD-3 measurement. We use the CMD-3
parameterization, i.e. an exponential saturated function to fit the $\sigma(\EE\to p\bar{p})$ data near threshold.
At higher energies, an exponential or a power-law damping function is used to parameterized the cross section.
Eq.~\ref{eq-pp1} and Eq.~\ref{eq-pp2} summarize the fit models to the $\sigma(\EE\to p\bar{p})$ data, where
$E_{thr}=2m_p$ is the mass threshold, $\sigma_{thr}\lesssim 1$~MeV is a variation scale parameter quoted
from CMD-3 measurement; $P_1$, $P_2$, $n$, and $E_0$ are free parameters. The fit results are shown in 
Fig.~\ref{xsec-pp}, where both fit models describe data reasonably well.

\begin{equation}
\sigma_{p\bar{p}}(\sqrt{s})=\left\{
	 	\begin{array}{lcl}
	 	P_1(1-e^{-\frac{\sqrt{s}-E_{thr}}{\sigma_{thr}}}) && {\sqrt{s}<E_0} \\
		\\
	 	P_1e^{-P_2(\sqrt{s}-E_0)} && {\sqrt{s}\geq E_0} \\
	 	\end{array} \right.
\label{eq-pp1}
\end{equation}
or
\begin{equation}
\sigma_{p\bar{p}}(\sqrt{s})=\left\{
	 	\begin{array}{lcl}
	 	P_1(1-e^{-\frac{\sqrt{s}-E_{thr}}{\sigma_{thr}}}) && {\sqrt{s}<E_0} \\
	 	\\
	 	P_1(\sqrt{s}-E_0+1)^{-n} && {\sqrt{s}\geq E_0} \\
	 	\end{array} \right.
\label{eq-pp2}
\end{equation}

\subsection{Fit the $\sigma(\EE\to\Lam)$ data}

Since the $\EE\to\Lam$ cross section shows a clear enhancement above the $\Lam$ 
threshold (c.f. Fig.~\ref{xsec-data}), Belle used a Breit-Wigner (BW)
resonance to fit the cross section data~\cite{belle-lamc}. However, the precise cross section measurement near 
threshold by the BESIII Collaboration shows a fast-rise structure near the $\Lam$ threshold~\cite{bes3-lamc}, 
which is extremely similar to the $\EE\to p\bar{p}$ process. If the cross section data both from Belle and BESIII
is considered, we find the original Belle model by only including a BW resonance simply fails in the fit. To overcome
these difficulties, we build a composite model by adding a continuum term to parameterized the $\EE\to\Lam$ 
cross section line shape in a wide energy range. Eq.~\ref{eq-lam} shows the fit probability-density-function (PDF),
\begin{equation}
\label{eq-lam}
\sigma_{\Lam}(\sqrt{s})=\frac{12\pi}{s}\frac{\Gamma_{e^+ e^-}}{\Gamma_{res}}\BR(R\rightarrow f)|A_{res}|^2+f(\sqrt{s}).
\end{equation}
where $A_{res}$ represents a resonance amplitude, and $f(\sqrt{s})$ represents a non-resonant continuum
contribution.

For the resonance amplitude, we use a standard $S$-wave relativistic-BW function (same as Belle)~\cite{PDG}
\begin{equation}
\label{eq-bw}
A_{res}=\frac{-\sqrt{s}\cdot\Gamma(\sqrt{s})}{s-M_{res}^2+i\sqrt{s}\cdot\Gamma(\sqrt{s})},
\end{equation}
where $\Gamma(\sqrt{s})$ is a $\sqrt{s}$-dependent resonance width, with
\begin{equation}
\label{eq-width}
\Gamma(\sqrt{s})=\Gamma_{res}\frac{q}{q_{res}}\frac{M_{res}}{\sqrt{s}}.
\end{equation}
Here $q$ is the $\Lambda_c^+$ momentum in the $\Lambda_c^+\Lambda_c^-$ center-of-mass (cm) frame, 
and $q_{res}$ is the corresponding momentum at $\sqrt{s}=M_{res}$.

For the continuum term, we borrow the model from the $\EE\to p\bar{p}$ cross section fit, i.e.
\begin{equation}
\label{eq-cont1}
f(\sqrt{s})=\left\{
		\begin{array}{lcl}
	 	P_1(1-e^{-\frac{\sqrt{s}-E_{thr}}{\sigma_{thr}}}) && {\sqrt{s}<E_0} \\
	 	\\
	 	P_2e^{-P_3(\sqrt{s}-E_0)}+P_4 && {\sqrt{s}\geq E_0} \\
	 	\end{array} \right.
\end{equation}
or
\begin{equation}
\label{eq-cont2}
f(\sqrt{s})=\left\{
		\begin{array}{lcl}
	 	P_1(1-e^{-\frac{\sqrt{s}-E_{thr}}{\sigma_{thr}}})&& {\sqrt{s}<E_0} \\
	 	\\
	 	P_2(\sqrt{s}-E_0+P_3)^{-n}+P_4 && {\sqrt{s}\geq E_0}. \\
	 	\end{array} \right.
\end{equation}
Similar to the $\EE\to p\bar{p}$ process, an exponential saturated function is used to describe the 
fast-rise structure near the $\EE\to\Lam$ production threshold, where $E_{thr}=2m_{\Lambda_c}$,
and $\sigma_{thr}=0.1$~MeV describing the variation scale and is fixed in the fit due to lack of
data near the $\Lam$ threshold. For the damping function at higher energy, we take the general form 
of an exponential function (Model-I) or a power-law function (Model-II) [$n=7$ is taken from the $\sigma(\EE\to p\bar{p})$ fit].
A constraint is also added to maintain the continuity at $E_0$ for $f(\sqrt{s})$.
$P_1$ ($=0$ for $\sqrt{s}<E_{thr}$), $P_2$, $P_3$, $P_4$ and $E_0$ are all floating parameters.

An binned maximum likelihood fit is performed to the Belle data and BESIII data simultaneously.
For the Belle data, since there are low statistics bins, a Poisson distribution is used to describe
the probability for observing $n_i^{\rm obs}$ events in each bin, i.e.
\begin{equation}
P_i=\frac{\mu_i^{n_i^{\rm obs}}e^{-\mu_i}}{n_i^{\rm obs}!}.
\end{equation}
Here
\begin{equation}
\mu_i=\lum^{\rm int}_i \sigma_i \eta_i^{\rm tot} (\sqrt{s})+n^{\rm bkg}_i
\end{equation}
where $\sigma_i=\frac{1}{\Delta m}\int_{m-\Delta m/2}^{m+\Delta m/2}\sigma(s)ds$ is the average 
cross section in the $i$-th bin (with bin width $\Delta m$); $\lum^{\rm int}_i=\int d\lum^{\rm int}_i$, $\eta_i^{\rm tot}$,
and $n^{\rm bkg}_i$ are the ISR luminosity, $\sqrt{s}$-dependent total efficiency (including branch fractions), and the estimated
background events in the $i$-th bin, respectively, which are quoted from Ref.~\cite{belle-lamc}.
For the BESIII data, each data set has enough statistics, and the cross section follows a Gaussian distribution
\begin{equation}
G_j=\frac{1}{\sqrt{2\pi}\delta}e^{\frac{-(\sigma_j-\sigma_{\Lam}(\sqrt{s}))^2}{2\delta^2}},
\end{equation}
where $\sigma_j$ is the measured cross section at cm energy $\sqrt{s}$, and $\delta$ is the uncertainty.

The likelihood function is written as $\lum=\prod_{i=1}^{42}P_i\prod_{j=1}^{4}G_j$, which is a combined
likelihood value calculated from both the Belle and BESIII data. In reality, we minimize $-2\ln\lum$ with the 
{\sc minuit} package in the CERN Program Library~\cite{minuit} to get the best estimation of the parameters.

\begin{figure}
\includegraphics[width=8cm]{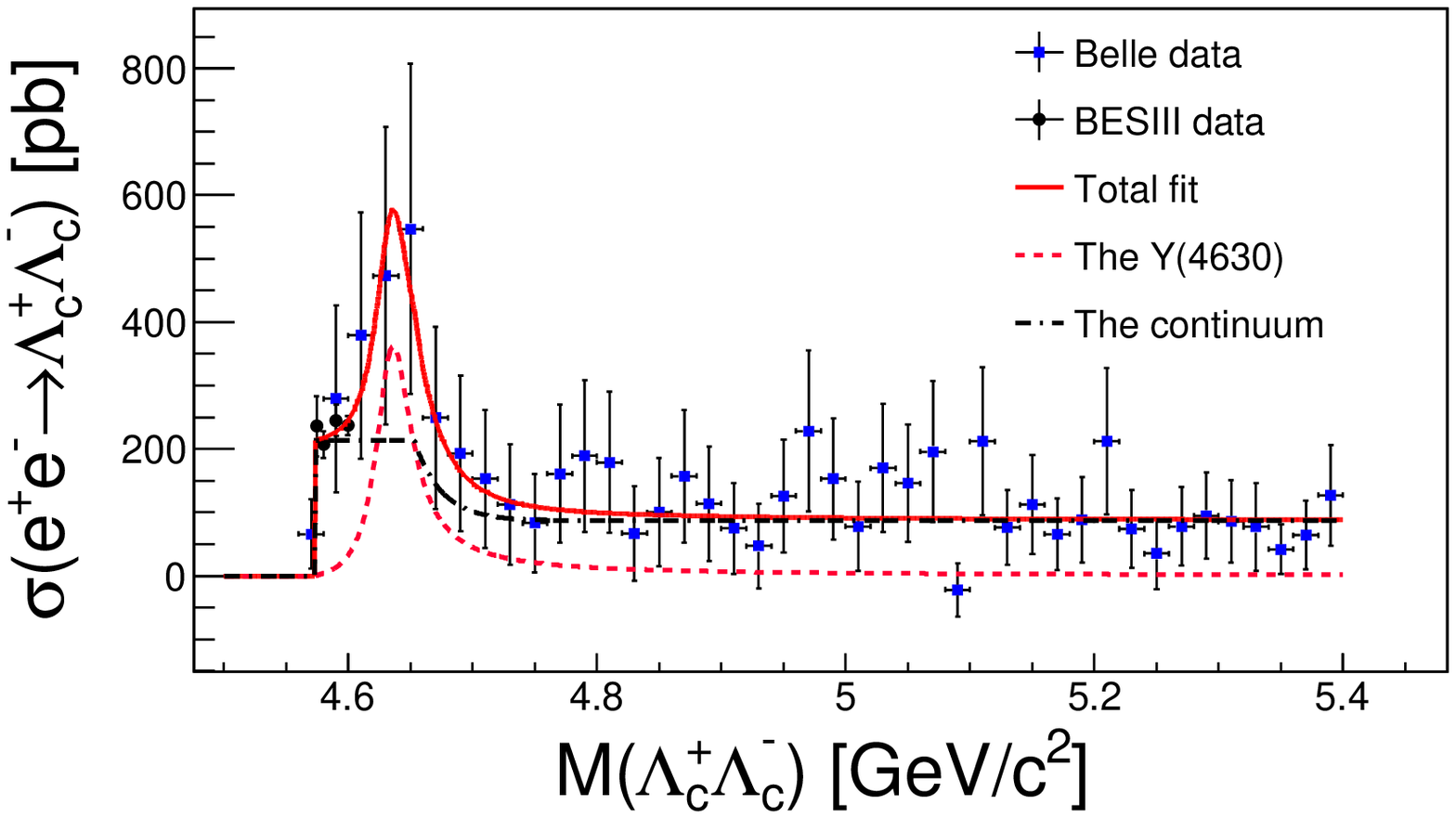}
\includegraphics[width=8cm]{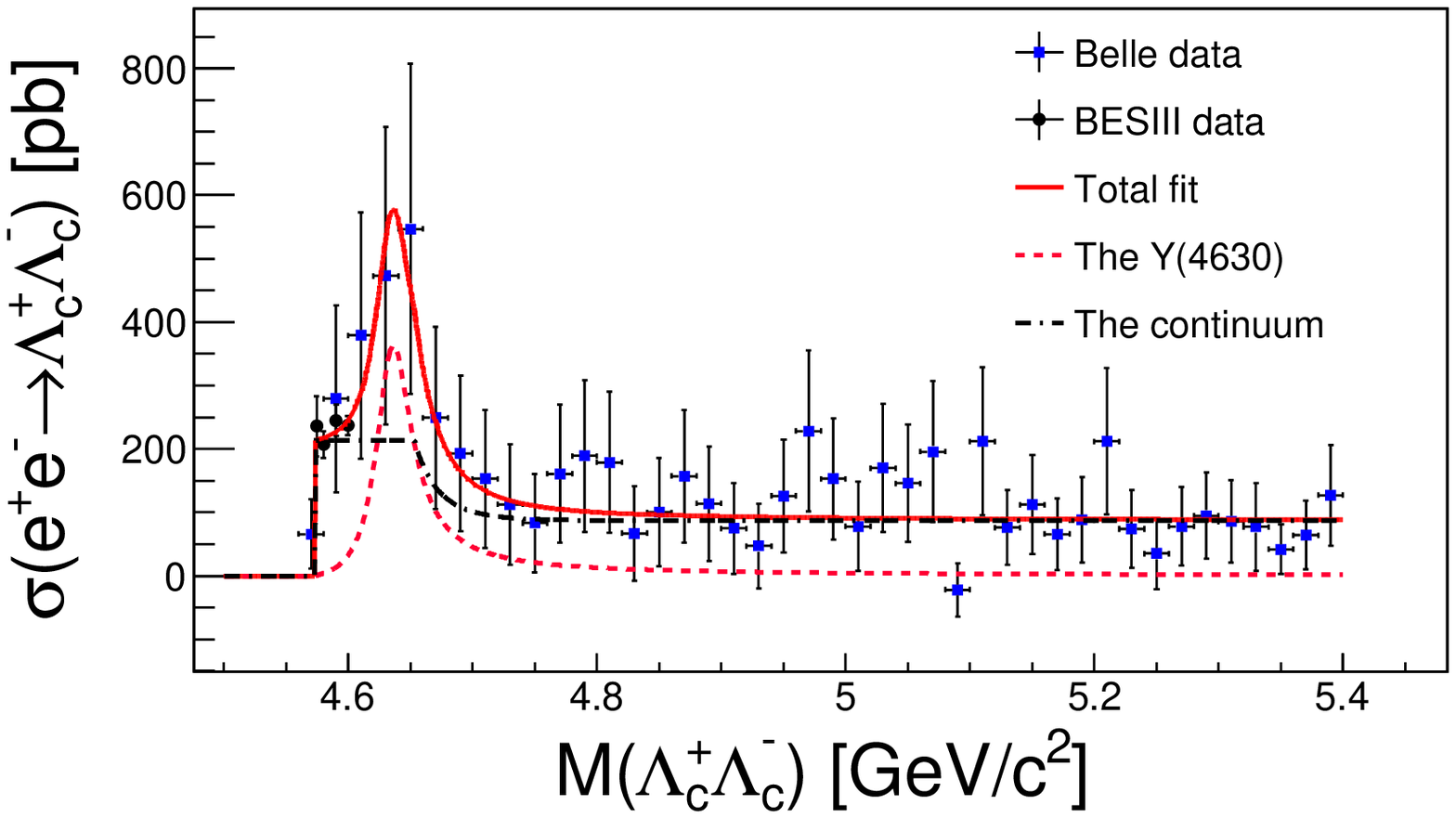}
\caption{A simultaneous binned likelihood fit to the $e^{+}e^{-}\rightarrow\Lambda_c^+\Lambda_c^-$ cross section
from both Belle (blue squares) and BESIII (black dots) measurements. 
The solid curves show the total fit results, the dashed curves show the BW resonance,
and the dotted-dashed curves show the continuum contribution with an exponential damping (left pannel) and
a power-law damping (right panel) at higher energy, respectively.}
\label{fit-xsec} 
\end{figure}

Figure~\ref{fit-xsec} shows the fit results, where good agreement can be seen between the fit
curves and the cross section data. A $\chi^2$-test is used to estimate the goodness of the fit,
which gives $\chi^2/ndf=0.96$ for Model-I (exponential damping) and $\chi^2/ndf=0.96$ for Model-II (power-law damping)
, respectively. The resonant parameters obtained from the fit are summarized in Table~\ref{para}.

\begin{table}
	\centering
	\caption{\label{para} The resonant parameters of $Y(4630)$ from the fits.	Here the errors are statistical only.}
	\begin{tabular}{ccc}
		\hline\hline
		~~Parameter~~ & ~~Model-I (Exponential damping)~~ & ~~Model-II (Power-law damping)~~ \\
		\hline
		$M[Y(4630)]$ & $4636.1_{-7.2}^{+9.8}~{\rm MeV}/c^2$ & $4636.3_{-7.3}^{+9.5}~{\rm MeV}/c^2$\\
		$\Gamma_{\rm tot}[Y(4630)]$ & $34.5_{-16.2}^{+21.0}~{\rm MeV}$ & $34.7_{-16.3}^{+21.0}~{\rm MeV}$\\
		$\Gamma_{e^+ e^-}\BR[Y(4630)\to\Lambda_c^+\Lambda_c^-]$ &  $18.3_{-6.1}^{+8.8}~{\rm eV}/c^2$ & $18.4_{-6.2}^{+8.5}~{\rm eV}/c^2$ \\
		\hline\hline
	\end{tabular}
\end{table}

\section{systematic errors}
\begin{figure}
	\begin{center}
		\includegraphics[width=12cm]{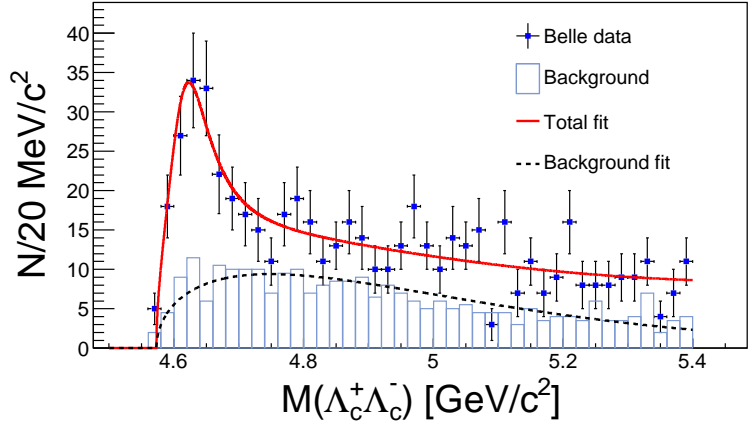}
		\caption{A binned maximum likelihood fit to the $M({\Lambda_c^+\Lambda_c^-})$ spectrum measured by Belle.}
		\label{repeat}
	\end{center}
\end{figure}

There are several systematic error sources which contribute to this analysis. The Belle measurement of
$\EE\to\Lam$ cross section used an ISR method, which produce a continuous $M(\Lam)$ mass spectrum~\cite{belle-lamc}.
When we take the cross section data from the Belle published measurement, only binned data is
provided. To estimate the effect of binning of the data, we repeat the Belle analysis with a binned maximum
likelihood fit, which is shown in Fig.~\ref{repeat}. Good agreement is observed between an un-binned fit
and a binned fit, and the small difference of the resonant parameters is taken as a systematic error 
due to the binning of data.

We build a composite model by including a continuum contribution in the PDF to describe the 
$\EE\to\Lam$ cross section. For the continuum term, there are two parameterization 
methods, i.e. at higher energy the cross section damps as an exponential or a power-law. We find
both these two parameterizations can describe data well, and the current data can 
not distinguish them. Thus, one of them is quoted as nominal result, and the difference of resonant 
parameters between these two models is taken as a systematic error due to the fit model.

A $\sqrt{s}$-dependent BW function is used to model a resonance in the fit PDF. Since the $Y(4630)$
is not so wide from our analysis, a constant full width BW function is also studied. The difference of
the resonant parameters is quoted as a systematic error due to the BW-parameterization.

In the fit, the non-$\Lam$ background shape is described by a smooth function 
$a\sqrt{M-M_{th}}\cdot e^{-(bM+cM^2)}$. Using the $\Lam$ mass sideband events directly to
represent the background in the fit, and the difference of the resonant parameters is estimated
as a systematic error due to backgrounds.

Table~\ref{sys-error} summarizes all the systematic sources and their contribution to the $Y(4630)$
parameters.

\begin{table}
	\centering
	\caption{\label{sys-error} The systematic error sources and their contribution to the $Y(4630)$ parameters.}
	\begin{tabular}{cccc}
		\hline\hline
		Source~~&~~$\delta M[Y(4630)]$~~&~~$\delta \Gamma[Y(4630)]$~~&~~$\delta \Gamma_{e^+ e^-}\BR[Y(4630)\to\Lambda_c^+\Lambda_c^-]$~\\
		\hline
		Binning of the data & $3.9~{\rm MeV}/c^2$ & $2.4~{\rm MeV}$ &$0.6~{\rm eV}/c^2$ \\
		Fit model & $0.2~{\rm MeV}/c^2$ &$0.2~{\rm MeV}$ & $0.1~{\rm eV}/c^2$ \\
		BW-paramaterization & $2.5~{\rm MeV}/c^2$ & $3.8~{\rm MeV}$ & $0.5~{\rm eV}/c^2$ \\
		Backgrounds & $6.5~{\rm MeV}/c^2$ & $3.4~{\rm MeV}$ & $0.8~{\rm eV}/c^2$ \\
		Total & $8.0~{\rm MeV}/c^2$&$5.6~{\rm MeV}$ & $1.1~{\rm eV}/c^2$ \\
		\hline\hline
	\end{tabular}
\end{table}

\section{Summary}

The $\EE\to\Lam$ cross section measured by BESIII recently shows a non-zero fast rise structure
near the threshold, which is different from a single BW resonance. To account for this feature,
we build a composite PDF to model the $\EE\to\Lam$ cross section by introducing a continuum
contribution, which is widely observed in $\EE\to p\bar{p}$ and $\Lambda\bar{\Lambda}$ processes.
A simultaneous likelihood fit is performed to both the Belle and BESIII data, and we obtain a mass
$M=[4636.1_{-7.2}^{+9.8} ($stat$)\pm 8.0($syst$)]$~MeV/$c^2$, a width 
$\Gamma_{\rm tot}=[34.5_{-16.2}^{+21.0} ($stat$)\pm 5.6($syst$)]$~MeV, and 
$\Gamma_{e^+ e^-}\BR[Y(4630)\to\Lambda_c^+\Lambda_c^-]=[18.3_{-6.1}^{+8.8} ($stat$)\pm 1.1 ($syst$)]$~eV/$c^2$,
respectively, for the $Y(4630)$ resonance. Compared with the previous Belle results, the width
of $Y(4630)$ is much more narrower in this study. Considering the recent observation of a
vector resonance $Y(4626)$ (mass $4625.9^{+6.2}_{-6.0}\pm 0.4$~MeV/$c^2$, and width
$49.8^{+13.9}_{-11.5}\pm 4.0$~MeV) in the $\EE\to D_s^+D_{s1}(2536)^-$ final state~\cite{y4626}, 
it strongly hints us $Y(4630)$ and $Y(4626)$ might be the same resonance.

About the nature of the continuum part, there is no clear answer yet. 
Normally, the one photon exchange model by only involving electro-magnetic interaction of 
$B\bar{B}$~\cite{xsec-bbar} shall work in low-energies where the wave-length of a photon is much larger 
than the size of a baryon, or in other words, the baryon is treated as a point-like particle. However,
in the $\EE\to\gamma^*\to p\bar{p}$, $\Lambda\bar{\Lambda}$, and $\Lam$ processes, the coupled photon
has an energy up to 4.6~GeV (wave length $\lambda \lesssim 0.3$~fm), which is comparable to 
(or smaller than) the baryon size. Thus, the baryon internal structure should be considered properly
in calculating the $\EE\to B\bar{B}$ cross sections. Further efforts are still needed to understand the
$\EE\to B\bar{B}$ cross sections near threshold.

Finally, the BESIII experiment successfully runs around 4.6~GeV in the past years, and in future
the beam energy can be upgraded to 2.45~GeV~\cite{whitepaper}. BESIII's future data set above 4.6~GeV
will greatly improve our knowledge about $\EE\to\Lam$, and also the vector resonance $Y(4630)$
and $Y(4626)$. Our study provides a reference for the future BESIII data taking for the study of
$\EE\to\Lam$. The Belle II experiment will accumulate 50~ab$^{-1}$ data in the following years~\cite{belle2white},
which also provides us valuable opportunities to precisely study $\EE\to\Lam$.

\acknowledgments
This work is supported in part by the National 1000-Talent Program for Young Professionals, and
by the National Natural Science Foundation of China under Contract No. 11975141.

\end{document}